\documentclass[letter,twocolumn]{jpsj2} 
\usepackage{bm}
 
\title{
Critical Slowing Down of Triangular Lattice Spin-3/2 Heisenberg 
\\Antiferromagnet Li$_7$RuO$_6$ via $^{7}$Li NMR 
}

\author{\textsc
{Yutaka ITOH}$^{1}$\thanks{E-mail: itoh@kuchem.kyoto-u.ac.jp}
, \textsc{Chishiro MICHIOKA}$^{1}$, \textsc{Kazuyoshi YOSHIMURA}$^{1}$,\\
\textsc{Kanako NAKAJIMA}$^{2}$, and 
 \textsc{Hirohiko SATO}$^{2}$
}
\inst{
$^{1}$Department of Chemistry, Graduate School of Science, Kyoto University, Kyoto 606-8502\\
$^{2}$Department of Physics, Chuo University, 1-13-27 Kasuga, Bunkyo-ku, Tokyo 112-8551
}

\abst{
We report $^{7}$Li NMR studies of single crystals of triangular-lattice Heisenberg antiferromagnet Li$_7$RuO$_6$. 
Slow critical divergence with a wide critical region of $\left|T/T_\mathrm{N} - 1\right|\leq$ 7 was observed in $^{7}$Li nuclear spin-lattice relaxation rate.
The slowing down of staggered spin fluctuations was analyzed in a renormalized classical region of a two-dimensional triangular-lattice non-linear sigma model. 
A spin stiffness constant was found to reduce to about 20 $\%$ from the value in a spin-wave approximation. 
The effect of spin frustration, e.g., Z$_2$ vortex excitations on the critical phenomena is suggested.  
}

\kword{Li$_7$RuO$_6$, triangular lattice, spin frustration, NMR, renormalized classical region}

\begin{document}
\maketitle

Spin frustration effect of a triangular lattice Heisenberg antiferromagnet has been one of the central issues in physics and chemistry of magnetic insulators. 
The ground state is classically a long range ordering state with the 120$^\circ$ spin structure, but
the quantum mechanical ground state might be a gapped or gapless quantum spin liquid~\cite{PWAnderson,Fradkin,Tsvelik}.  
Elementary excitations of Z$_2$ vortices, topologically stable point defects, are inherent even in classical triangular spin systems~\cite{KM1,KM2,MS,Mouhanna}. 
At finite temperatures, topological phase transition between the paramagnetic states is theoretically predicted from numerical simulations. 
 
Finite-temperature magnetic long range ordering of quasi low dimensional antiferromagnets may be ascribed to a three dimensional interaction and the anisotropy. 
The low dimensional characteristics are (i) the suppressed magnetic ordering temperature $T_\mathrm{N} < T_\mathrm{N}^{\mathrm{MF}}$ in the mean field approximation, (ii) the suppressed Curie-Weiss law, i.e., the maximum of uniform spin susceptibility at $T_\mathrm{max}$ due to short range ordering, and (iii) a wide critical region in two dimensions~\cite{Tsvelik}. 

Two dimensional square lattice Heisenberg antiferromagnet has been intensively studied 
through the studies of high-$T_{\rm c}$ cuprates~\cite{Fradkin,Tsvelik}. 
Our understanding a short range ordering has made rapid progress. 
In the renormalized classical region, an antiferromagnetic correlation length and a staggered spin susceptibility diverge exponentially toward $T_\mathrm{N}$ = 0 K~\cite{Chak1,Chak2}, 
which were observed in La$_2$CuO$_4$~\cite{Imai,Birg}.  
Such a wide critical region toward  $T_\mathrm{N}$ = 0 K is the notable feature of two dimensional square lattice Heisenberg antiferromagnets. 
Non-linear sigma model and field theoretical treatments turned out to be the relevant model and to give us powerful methods to describe the low energy excitations. 
They were also applied to the non-collinear frustrated magnetic systems like the triangular lattice~\cite{Azaria,Azaria1,Chuv1,Chuv2,Chuv3,Lecheminant}.
To our knowledge, however, the theoretical critical behavior has not been widely tested to the actual critical phenomena of triangular spin systems. 

\begin{figure}[bp]
	\begin{center} 
		    \includegraphics[width=8.5cm, clip]{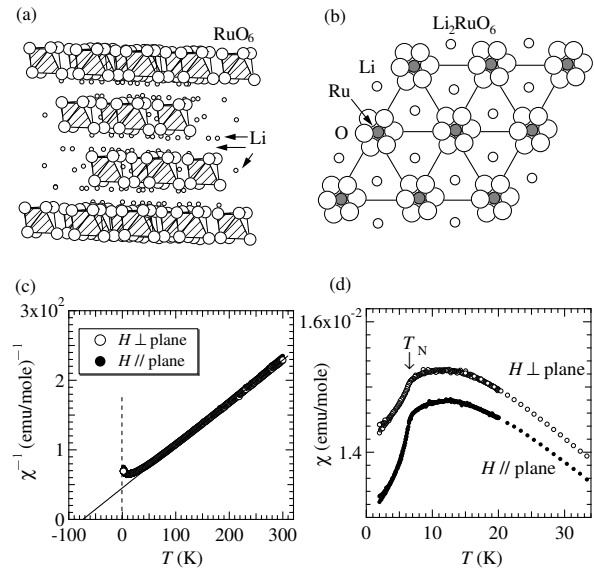}
	\end{center}
	\caption{(a) Schematics of crystal structure of Li$_{7}$RuO$_{6}$ and (b) the top view of a triangular lattice Ru plane. 
		(c) Inverse magnetic susceptibility $\chi^{-1}$ and (d) the magnetic susceptibility $\chi$ of a single crystal Li$_{7}$RuO$_{6}$. The solid lines are the best fits by an inverse Curie-Weiss law. 
		The arrow indicates $T_\mathrm{N}\approx$ 6.5 K. 
		}
		\label{fig_cry}
\end{figure}%
A delafossite-type Li$_7$RuO$_6$ has been at first reported as hexagonal ``Li$_{8-\delta}$RuO$_6$"~\cite{Synthesis1}.
However, a detailed analysis on powder~~\cite{Synthesis2} and on single crystal~\cite{Sato} revealed that the actual composition is ``Li$_7$RuO$_6$" and the structure is slightly distorted from an ideal hexagonal lattice.
Li$_7$RuO$_6$ is triclinic and has the superlattice structure of the Li deficiency in the double Li layers~\cite{Sato}. 
Figures \ref{fig_cry}(a) and \ref{fig_cry}(b) show the crystal structure and the triangular lattice, respectively. 
Single Li$_2$RuO$_6$ layer and double Li layers are stacked with each other. 
The Li$_2$RuO$_6$ layer consists  of a triangular lattice of RuO$_6$ octahedrons and a Li honeycomb lattice. 
The Li ions occupy 12 sites in the RuO$_6$ layer and 30 sites in the Li deficient layers in unit cell,
which are crystallographically inequivalent. 
For a hexagonal model of Li$_{8}$RuO$_6$, the staggered magnetic field from Ru would be cancelled out at the Ru-plane Li site. 
However, for the actual Li$_7$RuO$_6$, since the Li ions occupy non-ideal positions deviated from the center of the Ru triangle, then the 120$^\circ$ staggered spin fluctuations can be probed through the Li NMR. 
Li$_7$RuO$_6$ is a deformed triangular lattice system. 
The detail of the crystal growth and the structure analysis will be published in a separated paper~\cite{Sato}. 

Uniform magnetic susceptibility $\chi_{\alpha}$ ($\alpha$ = $\parallel$ and $\perp$ denote $H$ $\parallel$ plane and $H$ $\perp$ plane) shows a slightly anisotropic Curie-Weiss behavior $\chi$ = $C$/$(T - \Theta)$ with $C$ = 1.89 emu/mol$\cdot$Ru and $\Theta$ = $-$ 73 K at high temperatures. 
Figures \ref{fig_cry}(c) and \ref{fig_cry}(d) show the inverse magnetic susceptibility $\chi^{-1}$ up to 300 K and the magnetic susceptibility $\chi$ on an enlarged scale for a single crystal, respectively.
Upon cooling, $\chi_{\alpha}$ makes a maximum at about 12.5 K and
drops sharply at the Neel temperature $T_\mathrm{N}\approx$ 6.5 K~\cite{Synthesis2,Sato}.  
The broad maximum behavior indicates a low dimensional magnet. 
The Curie constant $C$ is close to the value for $S$ = 3/2 and $g$ = 2.  
For a Heisenberg spin Hamiltonian $\Sigma J_{nn}S_i\cdot S_j$ with the $z$ nearest-neighbor exchange interaction $J_{nn}$ between Ru ions, the Weiss temperature $\Theta$ is given by
\begin{equation}
\Theta = \frac{S(S+1)}{3}zJ_{nn},
\label{eq:HighT1}
\end{equation}
in the mean field approximation. 
Using $S$ = 3/2 (Ru$^{5+}$) and $z$ = 6, we estimated the superexchange interaction $J_{nn}$ = $-$ 9.7 K. 
Since $T_\mathrm{N}\approx$ 6.5 K and $\Theta$ = $-$73 K,    
we obtain the ratio of $T_\mathrm{N}/ |\Theta|\approx$ 0.089,
which is nearly the same as 0.082 for VCl$_2$ with $S$ = 3/2 (V$^{4+}$)~\cite{VCl2}.
These are the low dimensional characteristics of (i) and (ii). 
Thus, the layered compound Li$_7$RuO$_6$ is a quasi-two dimensional antiferromagnet. 

In this Letter, we report $^{7}$Li NMR studies of single crystals of triangular-lattice Heisenberg antiferromagnet Li$_7$RuO$_6$. We found slow critical divergence with a wide critical region of $\left|T/T_\mathrm{N} - 1\right|\leq$ 7 in $^{7}$Li nuclear spin-lattice relaxation rate, which is  due to the slowing down of staggered spin fluctuations. 
From the analysis  by renormalized classical fluctuations, we found the reduction of a spin stiffness constant, possibely due to the spin frustration effect.  

Single crystals of Li$_{7}$RuO$_{6}$ were grown from   
the mixture of RuO$_2$ and a large amount of LiCO$_3$ flux heated in an oxygen atmosphere at 950 $^{\circ}$C.
Typically 0.5$\times$0.5$\times$0.01 mm sized and plate-like single crystals were obtained~\cite{Sato}. 
The plane and the vertical axes are confirmed to be the $ab$ plane and the $c$ axis of the quasi-hexagonal lattice, respectively.    
X-ray diffraction patterns for the powdered samples indicated the samples in a single phase.
We performed $^{7}$Li (nuclear spin $I$ = 3/2 and nuclear gyromagnetic ratio $\gamma_{\rm n}/2\pi$ = 16.546 MHz/T) NMR spin-echo measurements at $H$ = 7.48414 T
for the samples in which the several single crystal pieces were put on a plate. 
The NMR frequency spectra were obtained from Fourier-transformed spin-echoes or summing them at several frequencies and frequency-swept spin-echo intensity. 
The $^{7}$Li nuclear spin-lattice relaxation curves were measured by an inversion recovery technique
and the relaxation times $T_1$'s were obtained from the fits by a stretched exponential function
of exp$\{$-($t$/$T_1$)$^\beta$$\}$ ($t$ is the time after the inversion pulse).   

\begin{figure}[tbp]
	\begin{center}
		\includegraphics[width=6.5cm, clip]{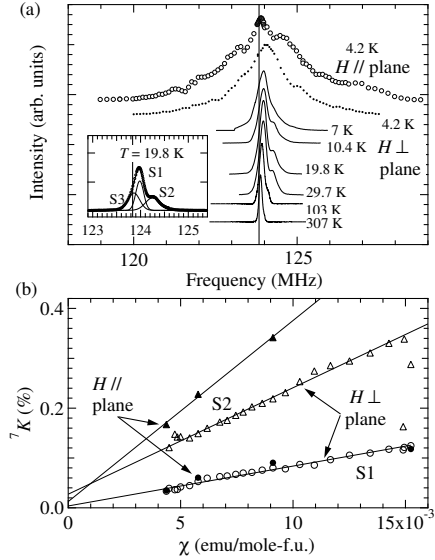}
	\end{center}
	\caption{(a) $^{7}$Li NMR frequency spectra of the single crystals of Li$_{7}$RuO$_{6}$
	at $H \perp$ plane (solid curves and closed circles) and at $H \parallel$ plane (open circles).
	The vertical line at 123.8343 MHz indicates the $^{7}$Li NMR spectrum peak of LiCl$aq$ for a reference of zero shift. 
	The inset shows the best fits by three Gaussian functions. 
	(b) Knight shift $^{7}K$ plotted against the bulk magnetic susceptibility $\chi_{\alpha}$ with temperature as an implicit parameter.
	}
		\label{fig_XT}
\end{figure}%
\begin{figure}[bp]
	\begin{center}
		\includegraphics[width=6.5cm, clip]{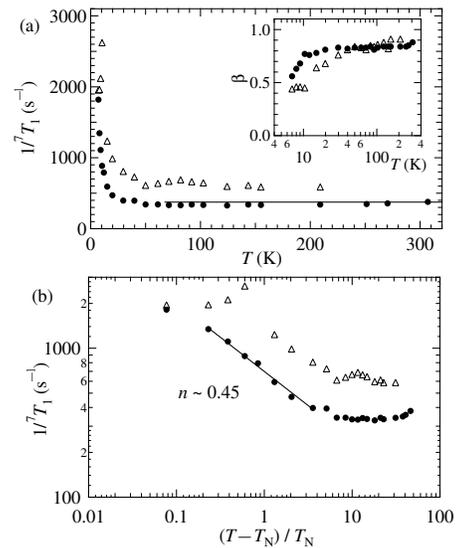}
	\end{center}
	\caption{
	(a) Temperature dependences of $^{7}$Li nuclear spin-lattice relaxation rates 1/$T_1$
	of S1 (closed circles) and S2 (open triangles) at $H \perp$ plane and $T > T_\mathrm{N}$.  
	The inset figure shows semi-log plots of the temperature dependences of the stretched exponents $\beta$. 
	(b) Log-log plots of  $^{7}$Li nuclear spin-lattice relaxation rates 1/$T_1$ at $H \perp$ plane
	against the reduced temperature $(T-T_\mathrm{N})/T_\mathrm{N}$.
	}
		\label{fig_T1}
\end{figure}%
Figure \ref{fig_XT}(a) shows $^{7}$Li NMR frequency spectra of the single crystals of Li$_{7}$RuO$_{6}$ at $H \perp$ plane (solid curves and closed circles) and at $H \parallel$ plane (open circles).
At $T >$ 200 K, a single sharp $^{7}$Li NMR spectrum and no quadrupole splits were observed. 
Upon cooling, the NMR spectrum shifts to higher frequency side and a weak signal separates and shifts more largely.
At 4.2 K, both NMR spectra at $H \perp$ plane and $H \parallel$ plane are broadened nearly symmetrically. 
The featureless symmetric broadening below $T_\mathrm{N}$ indicates the emergence of internal magnetic field of an incommensurate staggered moments along the $ab$ plane and the $c$ axis. 

The inset shows three Gaussian functions fit to the NMR spectrum at $H \perp$ plane and $T =$19.8 K. 
A sharp strong peak and a broad higher frequency peak are denoted by S1 and S2, respectively. 
S3 just adjusts the foot of the spectrum, whose Knight shift is $\sim$100 ppm and nearly independent of temperature. 
From the NMR intensity, 
the strong peak S1 and weak S2 can be assigned to the Li sites in the double Li layers and in the RuO$_6$ triangle lattice layer, respectively. 
The assignment at $H \parallel$ plane at $T > T_\mathrm{N}$ not shown in Fig.~\ref{fig_XT}(a) was less clear, since at least 4 peaks were observed. The lowest and highest frequency peaks are assigned to S1 and S2, resepctively. 

Figure \ref{fig_XT}(b) shows $^{7}$Li Knight shifts of S1 and S2 plotted against the bulk magnetic susceptibility $\chi_{\alpha}$ with temperature as an implicit parameter.
The bulk magnetic susceptibility $\chi_{\alpha}$ is expressed by the sum of a temperature dependent spin susceptibility $\chi_{\rm spin}$, the Van Vleck orbital susceptibility $\chi_{\rm VV}$, and
the diamagnetic susceptibility of inner core electrons $\chi_{\rm dia}$.
The $^{7}$Li Knight shift is expressed by $K_{\alpha}$ = $A_{\alpha}(i)\chi_{{\rm spin}, \alpha}/N_\mathrm{A}\mu_\mathrm{B}$+$K_\mathrm{dia}$ ($i$ =1 and 2 denote S1 and S2, $N_\mathrm{A}$ is Avogadro's number, and $\mu_\mathrm{B}$ is the Bohr magneton).
From the $K-\chi$ plots in Fig.~\ref{fig_XT}, we obtained positive hyperfine coupling constants
of $A_{\parallel}(1)\approx A_{\perp}(1) =$ 0.45 kOe/$\mu_\mathrm{B}$,
$A_{\parallel}(2) =$ 2.0 kOe/$\mu_\mathrm{B}$ and $A_{\perp}(2) =$ 1.2 kOe/$\mu_\mathrm{B}$. 
The positive $A(1)$ and $A(2)$ indicate the predominant role of a transferred hyperfine coupling.

Figure \ref{fig_T1}(a) shows temperature dependences of $^{7}$Li nuclear spin-lattice relaxation rates 1/$T_1$ of S1 and S2 at $H \perp$ plane and $T > T_\mathrm{N}$. 
The inset figure shows semi-log plots of the temperature dependences of the stretched exponents $\beta$. 
At high temperatures, 1/$T_1$ levels off, which indicates the exchange narrowing limit. 
At low temperatures, 1/$T_1$ shows the divergence behavior due to the critical slowing down of the staggered spin fluctuations.

In general, the high temperature limit of 1/$T_1$ in the exchange narrowing is expressed by~\cite{Moriya1}
\begin{equation}
\frac{1}{T_{1\infty}} = \sqrt{\frac{\pi}{2}}\frac{S(S+1)}{3}\frac{2\gamma_{\rm n}^2 A_{\parallel}^2}{\omega_{\rm ex}}
\label{eq:HighT1}
\end{equation}
and
\begin{equation}
\omega_{\rm ex}^2= \frac{2}{3}zS(S+1)\left(\frac{k_{\rm B}J_{nn}}{\hbar}\right)^2.
\label{eq:wex}
\end{equation}
The hyperfine field at the in-plane Li results from the 3 nearest neighbor Ru spins. 
The number of the nearest neighbor exchange coupled Ru spins is 6.    
Putting $z$ = 6,  $S$ = 3/2, $A_{\parallel}(2) = $ 2.0 kOe/$\mu_\mathrm{B}$ and $J_{nn}$ = $-$9.7 K for eqs.~(\ref{eq:HighT1}) and (\ref{eq:wex}), we obtain 
1/$T_{1\infty}$ = 340 s$^{-1}$, which is the same order of magnitude but slightly smaller than the actual 1/$T_{1}\approx$ 400 and 600 s$^{-1}$ for S1 and S2 above 200 K, respectively. 
The experimental value of $T_{1}$ depends on the fitting function more or less.
Using the fitting function of a double exponential function, we obtained a long component of 1/$T_{1}\approx$ 277 and 435 s$^{-1}$ for S1 and S2, respectively,
being the same order of magnitude of the estimated 1/$T_{1\infty}$. 

Figure \ref{fig_T1}(b) shows log-log plots of  $^{7}$(1/$T_1$)  against the reduced temperature $(T-T_\mathrm{N})/T_\mathrm{N}$. 
The critical divergence of 1/$T_{1}$ starts from  $\left|T/T_\mathrm{N} - 1\right |$ $\sim$ 7.
Although such a wide critical region is not usually regarded as a three dimensional critical region,
we tried to apply a power law of $\left|T_\mathrm{N}/(T - T_\mathrm{N})\right|^n$ and then obtained $n$ = 0.45.  
This is close to the mean field value~\cite{Moriya3}. 
\begin{figure}[tbp]
	\begin{center}
		\includegraphics[width=7.5cm, clip]{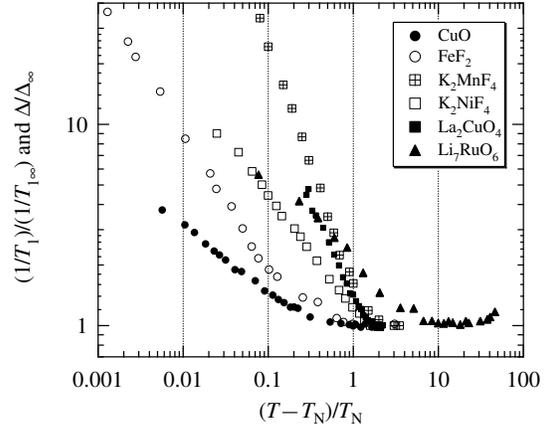}
	\end{center}
	\caption{Log-log plots of the nuclear spin-lattice relaxation rate (1/$T_1$)/(1/$T_{1\infty}$) or the continuous-wave (cw) NMR linewidth $\Delta/\Delta_{\infty}$ against the reduced temperature $(T-T_\mathrm{N})/T_\mathrm{N}$ for the critical phenomena of CuO~\cite{CuO}, FeF$_2$~\cite{FeF2}. K$_2$MnF$_4$~\cite{K2MnF4}, K$_2$NiF$_4$~\cite{K2NiF4}, La$_2$CuO$_4$~\cite{Imai}, and Li$_7$RuO$_6$ (S1).  
	The circles, squares and triangles are three dimensional magnets, quasi two dimensional square lattices, and a triangular lattice, respectively. The solid and open symbols are the data by spin-echo and continuous-wave (cw) measurements, respectively.
	}
		\label{fig_CR}
\end{figure}%

Figure \ref{fig_CR} shows log-log plots of the nuclear spin-lattice relaxation rate (1/$T_1$)/(1/$T_{1\infty}$) or the continuous-wave (cw) NMR linewidth $\Delta/\Delta_{\infty}$ against the reduced temperature $\epsilon\equiv$$(T-T_\mathrm{N})/T_\mathrm{N}$ for the critical phenomena of CuO~\cite{CuO}, FeF$_2$~\cite{FeF2}, K$_2$MnF$_4$~\cite{K2MnF4}, K$_2$NiF$_4$~\cite{K2NiF4}, La$_2$CuO$_4$~\cite{Imai}, and Li$_7$RuO$_6$.  
The onset temperatures of the increase in $T_{1\infty}$/$T_1$ and linewidth $\Delta/\Delta_{\infty}$ 
are the beginning of the critical slowing down of relevant spin fluctuations
and indicate the width of the individual critical region. 
For a narrow critical region of $\left|\epsilon\right|\leq$ 0.1 of CuO and FeF$_2$, the critical exponent of three dimensional critical slowing down was estimated~\cite{CuO,FeF2}. 
The wide critical region of  $\left|\epsilon\right|\leq$ 1.0 of La$_2$CuO$_4$
was successfully understood by a two dimensional short range ordering effect in the renormalized classical region of the square lattice~\cite{Imai}. 
The wide critical regions  of square lattices K$_2$MnF$_4$ and K$_2$NiF$_4$ were also understood by the renormalized classical fluctuations~\cite{ItohX,Birg2}.
The critical region of $\left|\epsilon\right|\leq$ 7 of Li$_7$RuO$_6$ is wider than those of the two dimensional square lattice spin systems.    
One should note the another striking feature of the slow divergence of 1/$T_1$ of Li$_7$RuO$_6$. 
The wide but slow critical divergence of 1/$T_1$ characterizes the critical phenomenon of Li$_7$RuO$_6$. 
   
For the frustrated quantum antiferromagnets, the non-linear sigma model description tells us the magnetic correlation length~\cite{Azaria,Chuv1,Chuv2}
\begin{equation}
\xi\propto\frac{1}{\sqrt{T}} \mathrm{exp}(4\pi\rho_{\alpha}/T)
\label{eq:xiSU2}
\end{equation}
with a spin stiffness constant $\rho_{\alpha}$ ($\alpha$ = $\perp$ and $\parallel$ plane) and 
the nuclear spin-lattice relaxation rate~\cite{Chuv1,Chuv2}
\begin{eqnarray}
\frac{1}{T_1}\propto T^3\mathrm{exp}(4\pi\rho_{\alpha}/T). 
\label{eq:T1SU2} 
\end{eqnarray}  
The spin stiffness constant $\rho_{\perp}$ is given by
\begin{equation}
\rho_{\perp} = \frac{\sqrt 3}{2}Z_{\perp}S^2|J| \approx 1.51|J|,
\label{eq:RGZ}
\end{equation}
where a renormalization factor $Z_{\perp}$ is estimated by a spin-wave approximation and 1/$S$ expansion~\cite{Chuv3,Lecheminant}.   
Equations (\ref{eq:xiSU2}) and (\ref{eq:T1SU2}) are applicable to the low temperature states at $T$ $\ll$ 2$\pi\rho_{\alpha}$~\cite{Chuv1,Chuv2}. 
 
The NMR linewidth can serve as a probe of the magnetic correlation length $\xi$ in real space. 
The linewidth of the S2 NMR spectrum in Fig. \ref{fig_XT}(a) increased rapidly at 70 K upon cooling,
which did not scale with the Knight shift $K$ but was similar to the divergence in 1/$T_1$.  
Thus, we may assume the temperature dependent $\xi$ as in eq.~(\ref{eq:xiSU2}). 
\begin{figure}[tbp]
	\begin{center}
		\includegraphics[width=6.7cm, clip]{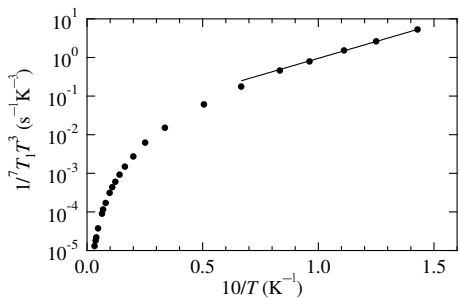}
	\end{center}
	\caption{Semi-logarithmic 1/$T_1T^3$ against 10/$T$ for Li$_7$RuO$_6$ (S1) at $H \perp$ plane.
	The straight line is the fitting function of eq.~(\ref{eq:T1SU2}).
	}
		\label{fig_RC}
\end{figure}%

Figure \ref{fig_RC} shows semi-logarithmic 1/$T_1T^3$ against 10/$T$ for Li$_7$RuO$_6$ at $H \perp$ plane. 
The straight line is the fitting result of eq.~(\ref{eq:T1SU2}).
From the slope of the line at lower temperatures $T <$ 15 K, 
we estimated the exchange interaction $\left|J\right|$ = 2.1 K. 
Since 2$\pi\rho_{\perp} =$ 20 K, the fit range is justified {\it a posteriori}.   
The interaction $\left|J_{nn}\right|$ = 9.7 K from the Weiss temperature $\Theta$ reduces 
to $\left|J\right|$ = 2.1 K at low temperatures. 
Since $J_{nn}$ should not strongly depend on temperature in the range of 4.2 K $< T <$ 300 K, 
the alternative reduction should trace back to the renormalization factor $Z_{\perp}$
and the stiffness constant $\rho_{\perp}$.
$Z_{\perp}$ should reduce to about 20 $\%$ from the value in the spin-wave approximation.   
This reduction might be the effect of the spin frustration, e.g., the topological $Z_2$ vortex excitations on the spin correlation, which agrees with the numerical simulations~\cite{Mouhanna}.  
The reduced $Z_{\perp}$ and $\rho_{\perp}$ are just to put the slow divergence of 1/$T_1$ in another way.      

The wide critical region has been observed for the other $S$ = 3/2 triangular lattice systems such as VCl$_2$~\cite{VCl2ND}, HCrO$_2$~\cite{Ajiro}, and LiCrO$_2$~\cite{Ajiro,Mahajan}.
It is, however, less clear whether the slow critical divergence is ubiquitous in the triangular lattices. 

In conclusion, we found slow critical divergence with a wide critical region of $\left|T/T_\mathrm{N} - 1\right|\leq$ 7 in $^{7}$Li nuclear spin-lattice relaxation rate for the triangular-lattice antiferromagnet Li$_7$RuO$_6$.   
We applied renormalized classical staggered spin fluctuations to the slow divergence
and then obtained the reduction of a spin stiffness constant, suggesting the spin frustration effect.  
 
We thank D. Mouhanna for valuable discussions. 
This work was supported in part by a Grant-in-Aid for Science Research on
Priority Area, ``Invention of Anomalous Quantum Materials," from the Ministry
of Education, Culture, Sports, Science and Technology of Japan (Grant No. 16076210)
and in part by a Grant-in-Aid for Scientific Research from the Japan Society
for the Promotion of Science (Grant No. 19350030).

\end{document}